\documentclass[aps,pra,twocolumn,superscriptaddress,floatfix,nofootinbib]{revtex4-2}

\usepackage[utf8]{inputenc}
\usepackage[T1]{fontenc}
\usepackage[english]{babel}
\usepackage{amsmath,amssymb,amsfonts,bm}
\usepackage{graphicx}
\usepackage{booktabs}
\usepackage{xcolor}
\usepackage{xspace}
\usepackage{hyperref}

\hypersetup{colorlinks=true,linkcolor=blue,citecolor=blue,urlcolor=blue}

\newcommand{\AD}{\ensuremath{\mathrm{AD}}\xspace}
\newcommand{\DEP}{\ensuremath{\mathrm{DEP}}\xspace}
\newcommand{\DPH}{\ensuremath{\mathrm{DPH}}\xspace}
\newcommand{\ch}{\ensuremath{\mathrm{ch}}\xspace}

\newcommand{\Tr}{\ensuremath{\mathrm{Tr}}}
\newcommand{\diag}{\ensuremath{\mathrm{diag}}}
\newcommand{\id}{\mathbb{I}}

\newcommand{\mcP}{\mathcal{P}}

\begin{document}

\title{Taking Advantage of Noise in Distributed Random Quantum Circuits}

\author{J. Montes}
\affiliation{Grupo de Sistemas Complejos, ETSIME, Universidad Polit\'ecnica de Madrid,
R\'ios Rosas 21, 28003 Madrid, Spain}

\author{F. Borondo}
\affiliation{Departamento de Química, Universidad Autónoma de Madrid, Cantoblanco, 28049 Madrid, Spain}

\author{Gabriel G. Carlo}
\affiliation{Comisión Nacional de Energía Atómica, CONICET, Departamento de Física, Av.\ del Libertador 8250, 1429 Buenos Aires, Argentina}

\date{\today}

\begin{abstract}
Adding noise can make a random quantum circuit look faster without making its unitary dynamics more random. This distinction is especially relevant in modular processors, where local gates randomize each core and scarce inter-core communication must spread that randomness across the full device. In this paper, we study this problem with a reduced second-moment transfer-matrix theory for Pauli second moments in distributed random circuits affected by the amplitude-damping, depolarizing, and dephasing noise channels. The key step is to resolve the noisy spectrum into two branches: a radial branch, describing dissipative loss of non-identity Pauli weight, and an angular branch, describing Haar-like mixing within the surviving nontrivial sector. This separation gives a simple weak-noise criterion: noise is useful for angular randomization only when it suppresses the longitudinal Bloch component more strongly than the transverse plane. Among the three channels considered, this selects amplitude damping as the only locally favorable case, while depolarizing noise is neutral and dephasing is dominated by radial loss. For multicore architectures, we derive a universal first-order law for radial leakage and track the angular branch numerically across different channels, topologies, and core partitions. The results reveal narrow windows of genuine noise-assisted Haar mixing, most clearly for amplitude damping, but rule out a generic speed-up by noise. The framework therefore distinguishes useful noisy randomization from mere dissipation.
\end{abstract}

\maketitle

\section{Introduction}
\label{sec:introduction}

Large-scale quantum computation will likely require architectures in which several quantum modules, processors, or cores are connected through comparatively limited communication links. This design principle is becoming relevant across superconducting, photonic, neutral-atom, and modular platforms, where the cost of moving quantum information between distant regions can be very different from the cost of applying local gates~\cite{Madsen2022,Bluvstein2024,Rad2025,Jnane2022,Hetenyi2024}. As a result, compilation, routing, and inter-core scheduling are no longer secondary engineering details, but part of the physical architecture of the computation itself~\cite{Wu2024,Yam2025,Vazquez2024,Wu2025,Dalton2025,Jeng2025,AndresMartinez2024,DeBone2024,AndresMartinezHeunen2019,Rached2025}. At the same time, present devices remain noisy, even though recent progress in quantum hardware, logical memories, and error-corrected demonstrations continues to push the field toward fault-tolerant operation~\cite{Preskill2018,Arute2019,Pan2022,King2025,GoogleQuantumAI2025,Bravyi2024,Bland2025}. The broader idea that noise can sometimes be exploited, rather than only mitigated, has also appeared in algorithms, including the dynamical mean-field theory (DMFT) approach to strongly correlated systems~\cite{Bertrand2025} and robust ground-state energy estimation based on quantum exponential least squares under depolarizing noise~\cite{Ding2026}, as well as in quantum reservoir computing~\cite{Domingo2022,Domingo2023}. Related work by Cirac and collaborators has studied stability to errors and quantum advantage in noisy analog quantum simulation, and benchmarked ground-state preparation algorithms in the presence of noise~\cite{Trivedi2024,Kashyap2025,Molpeceres2026}.

Random circuits provide a natural way to study this regime. Their convergence toward Haar statistics can be characterized through low-order moments, and the second moment already contains much of the information needed to describe scrambling, anticoncentration, and the formation of unitary designs~\cite{HarrowLow2009,Brandao2016,BrownViola2010,Haferkamp2022}. For unitary random circuits, the evolution of these second moments in the Pauli representation is governed by a Markov matrix, whose spectral gap gives a compact measure of the rate at which the circuit approaches the corresponding Haar moment~\cite{Weinstein2008}. In a previous noiseless study of distributed random circuits~\cite{MontesPRA2026}, this approach revealed a simple architectural trade-off: local gates are needed to randomize each core, but communication gates are needed to distribute that randomness globally. The competition between these two processes produces an optimal number of intra-core layers before communication.

The question addressed here is what remains of that optimum when local noise is present. Notice that this is not a purely quantitative modification of the noiseless problem. In a unitary circuit, the non-identity Pauli sector is conserved, and the relevant spectral gap measures mixing within that sector. In an open circuit, this conservation is lost. A noisy channel can suppress nontrivial Pauli components, push weight toward the identity, and increase an apparent spectral gap simply by dissipating information. Therefore, a larger noisy gap does not automatically mean faster Haar-like randomization. It may instead reflect the loss of the degrees of freedom whose angular distribution one wanted to randomize in the first place.

This distinction is the central point of the paper. We introduce a sector-resolved transfer-matrix description for distributed random circuits with three standard one-qubit noise channels: amplitude damping (\AD), depolarizing noise (\DEP), and dephasing (\DPH)~\cite{NielsenChuang2010,Mangini2022}. The noisy second-moment dynamics separates into a radial sector, associated with the decay of the total non-identity Pauli weight, and an angular sector, associated with redistribution inside the surviving nontrivial Pauli components. This separation allows us to distinguish genuine Haar-like mixing from purely dissipative contraction.

The main analytical result is a weak-noise criterion for when local noise can be favorable for angular mixing. The criterion depends only on whether the homogeneous part of the channel contracts longitudinal second moments more strongly or more weakly than transverse second moments before the random local rotation. We use this criterion as an organizing tool for the channel-by-channel comparison below, where the angular effect of each noise model is read together with the radial leakage it produces.

At the distributed level, the two sectors behave differently. The radial sector admits a universal first-order law that follows from counting Pauli strings and does not require diagonalizing the full multicore transfer matrix. The angular sector, by contrast, remains architecture-dependent: it depends on the communication graph, the number of qubits per core, the number of local layers before communication, and the local random-gate ensemble. This is precisely the sector that contains the physically relevant information about Haar-like mixing, and it must therefore be followed separately from the dissipative branch.

The numerical results support this picture while also placing clear limits on it. The raw angular gaps increase for amplitude damping, depolarizing noise, and dephasing, but this increase is not by itself a reliable signature of useful noise. Once radial decay is factored out, genuine angular improvement is confined to narrow weak-noise windows, with the most consistent enhancement occurring for amplitude damping. The conclusion is therefore not that noise generically helps random circuits. Rather, the result is a diagnostic criterion: it identifies when a noisy spectral speed-up corresponds to useful angular mixing and when it is only the spectral footprint of dissipation.

The paper is organized as follows. Section~\ref{sec:model} introduces the reduced second-moment transfer description of distributed random circuits, fixes the notation for the noiseless multicore architecture, and derives the one-qubit noisy transfer blocks for amplitude damping, depolarizing noise, and dephasing. Section~\ref{sec:radial_angular} develops the radial--angular decomposition, the weak-noise criterion, the universal radial law for the multicore circuit, and the branch-resolved diagnostics used below. Section~\ref{sec:results} presents the numerical results for the different channels, topologies, and core partitions. Finally, Section~\ref{sec:discussion_conclusions} summarizes the physical consequences and the limitations of noise-assisted Haar-like mixing in distributed quantum circuits. The appendix gives the analogous reduced derivation for depolarizing noise and dephasing, keeping the main text focused on the nonunital amplitude-damping case.

\section{Reduced transfer-matrix description of a noisy distributed circuit}
\label{sec:model}

\subsection{Distributed architecture and second moments}

We consider $N=Cq$ qubits divided into $C$ identical cores of $q$ qubits. One circuit cycle consists of $k$ intra-core layers followed by one inter-core communication layer associated with a graph $G$. If $E_G$ is the number of graph edges, the elementary gate count assigned to one cycle is
\begin{equation}
D=Ck+E_G,
\label{eq:D_definition}
\end{equation}
and the communication ratio is
\begin{equation}
r=\frac{E_G}{Ck}.
\label{eq:r_definition}
\end{equation}
The intra-core layer contains one-qubit gates with probability $p_1$ and two-qubit intra-core gates with probability $p_2=1-p_1$. The inter-core operations are unitary two-qubit gates. Their detailed action matters for angular mixing, but not for the first-order radial law derived below.

Let $\mcP_N$ be the $N$-qubit Pauli group modulo phases. A density matrix can be written as
\begin{equation}
\omega=\frac{1}{2^N}\sum_{P\in\mcP_N}r_P P,
\qquad
r_P=\Tr(P\omega).
\label{eq:pauli_expansion}
\end{equation}
For a random circuit ensemble, the second moments
\begin{equation}
m_P=\mathbb{E}(r_P^2)
\end{equation}
evolve linearly,
\begin{equation}
m'_P=\sum_{Q\in\mcP_N}M_{PQ}m_Q.
\label{eq:second_moment_full}
\end{equation}
For unitary gates, $M$ is a Markov matrix whose entries are averages of squared Pauli-transfer coefficients. When noise is included, we keep the same linear second-moment construction, but the resulting objects are transfer matrices rather than, in general, stochastic Markov matrices. Following Refs.~\cite{Weinstein2008,MontesPRA2026}, we reduce the $4^N$-dimensional Pauli space by identifying the two transverse components $X$ and $Y$ on each qubit. Each local symbol is therefore
\begin{equation}
I,\qquad Z,\qquad \perp,
\end{equation}
and the reduced moment vector has dimension $3^N$. This reduction preserves the distinction between longitudinal and transverse second moments, which is precisely the distinction needed to separate the effects of different noise channels.

\subsection{Noiseless local mixing block}

For one qubit,
\begin{equation}
\omega=\frac{1}{2}(\sigma_0+c_x\sigma_x+c_y\sigma_y+c_z\sigma_z).
\label{eq:single_qubit_bloch}
\end{equation}
A unitary $U\in SU(2)$ induces a rotation $O(U)\in SO(3)$ of the Bloch vector. We assume an ensemble with vanishing cross averages and transverse symmetry. Its second moments are parametrized by a single number $c$:
\begin{align}
\mathbb{E}(O_{zz}^2)&=c,\\
\mathbb{E}(O_{xz}^2)=\mathbb{E}(O_{yz}^2)&=\frac{1-c}{2},\\
\mathbb{E}(O_{zx}^2)=\mathbb{E}(O_{zy}^2)&=\frac{1-c}{2},\\
\mathbb{E}(O_{xx}^2)=\mathbb{E}(O_{yy}^2)&=\frac{1+c}{4},\\
\mathbb{E}(O_{xy}^2)=\mathbb{E}(O_{yx}^2)&=\frac{1+c}{4}.
\end{align}
With
\begin{equation}
p_0=\mathbb{E}(c_0^2),\qquad
p_z=\mathbb{E}(c_z^2),\qquad
p_\perp=\mathbb{E}(c_x^2+c_y^2),
\end{equation}
where $c_0=\Tr(\sigma_0\omega)=1$,
the reduced one-qubit block is
\begin{equation}
R_0(c)=
\begin{pmatrix}
1&0&0\\
0&c&\frac{1-c}{2}\\
0&1-c&\frac{1+c}{2}
\end{pmatrix}.
\label{eq:R0}
\end{equation}
The transient $2\times2$ block has two eigenvalues,
\begin{equation}
\lambda_R(0)=1,\qquad
\lambda_A(0)=\mu_A(c)=\frac{3c-1}{2}.
\label{eq:noiseless_eigenvalues}
\end{equation}
The eigenvector of $\lambda_R=1$ is proportional to $(1,2)^T$ and represents an isotropic non-identity second moment: one longitudinal and two transverse directions. The angular eigenvector can be chosen as $(1,-1)^T$ and measures the imbalance between the $z$ axis and the transverse plane.

The terms radial and angular refer to this Bloch-sphere geometry. The quantity
\[
p_z+p_\perp=\mathbb{E}(c_x^2+c_y^2+c_z^2)
\]
is the mean squared Bloch radius, or equivalently the total non-identity Pauli weight of the one-qubit second moment. A perturbation along $v_R=(1,2)^T$ changes this weight while keeping the isotropic ratio between the longitudinal direction and the two transverse directions. We therefore call this the radial mode. A perturbation along $v_A=(1,-1)^T$ has zero net contribution to $p_z+p_\perp$; it transfers second-moment weight between the $z$ direction and the transverse plane. In the Bloch picture it changes the angular distribution on a sphere of fixed radius, and we call it the angular mode.

For a full multicore cycle, the effective gap per elementary gate is
\begin{equation}
\Delta(k)=1-|\lambda(k)|^{1/D},
\qquad
D=Ck+E_G.
\label{eq:gapdef}
\end{equation}
This normalization is essential: it compares different local depths at fixed cost per elementary operation.

\subsection{Noisy one-qubit blocks}

The noisy local step is taken to be
\begin{equation}
\omega'=U\,\Lambda_{\ch,\gamma}(\omega)\,U^\dagger,
\label{eq:noise_then_rotation}
\end{equation}
where $\ch=\AD,\DEP,\DPH$ labels the channel. In Bloch form,
\begin{equation}
\bm c^{\,\mathrm{noise}}=A_{\ch}(\gamma)\bm c+\bm a_{\ch}(\gamma),
\qquad
\bm c'=O(U)\bm c^{\,\mathrm{noise}}.
\end{equation}
We keep the same second-moment reduction as in the unitary case. Cross terms are removed by the random rotation ensemble and by the reduced moment closure.

For the \AD channel,
\begin{equation}
K_0=
\begin{pmatrix}
1&0\\
0&\sqrt{1-\gamma}
\end{pmatrix},
\qquad
K_1=
\begin{pmatrix}
0&\sqrt{\gamma}\\
0&0
\end{pmatrix},
\end{equation}
In the Schrödinger picture, the noisy part of the step is
\begin{equation}
\Lambda_{\AD,\gamma}(\omega)
=K_0\omega K_0^\dagger+K_1\omega K_1^\dagger .
\end{equation}
This gives
\begin{equation}
\begin{pmatrix}
c_0\\
c_x\\
c_y\\
c_z
\end{pmatrix}
\mapsto
\begin{pmatrix}
c_0\\
\sqrt{1-\gamma}\,c_x\\
\sqrt{1-\gamma}\,c_y\\
(1-\gamma)c_z+\gamma
\end{pmatrix}.
\end{equation}
The reduced second-moment block follows by keeping the identity component explicitly during the nonunital part of the channel. Define
\begin{equation}
\tilde{\bm c}=(c_0,c_x,c_y,c_z)^T,\qquad c_0=1,
\end{equation}
At this stage we keep the four second moments
$p_\mu=\mathbb{E}(c_\mu^2)$, $\mu\in\{0,x,y,z\}$,
before imposing the transverse reduction. We then write the \AD action as a linear map on this four-component vector,
\begin{equation}
A_{\AD}(\gamma)=
\begin{pmatrix}
1&0&0&0\\
0&\sqrt{s}&0&0\\
0&0&\sqrt{s}&0\\
\gamma&0&0&s
\end{pmatrix},
\qquad s=1-\gamma.
\label{eq:AAD_four_component}
\end{equation}
The subsequent rotation is $L(U)=1\oplus O(U)$, so the one-step Bloch transfer matrix is $T_{\AD}(U,\gamma)=L(U)A_{\AD}(\gamma)$. In the ordered basis $(0,x,y,z)$,
\begin{equation}
T_{\AD}(U,\gamma)=
\begin{pmatrix}
1&0&0&0\\
\gamma O_{xz}&\sqrt{s}O_{xx}&\sqrt{s}O_{xy}&sO_{xz}\\
\gamma O_{yz}&\sqrt{s}O_{yx}&\sqrt{s}O_{yy}&sO_{yz}\\
\gamma O_{zz}&\sqrt{s}O_{zx}&\sqrt{s}O_{zy}&sO_{zz}
\end{pmatrix}.
\label{eq:TAD_four_component}
\end{equation}
The transfer matrix for second moments is obtained by squaring these entries and averaging over the rotation ensemble.
Assuming the reduced closure $\mathbb{E}(c_\mu c_\nu)=0$ for $\mu\neq\nu$, the second moments obey
\begin{equation}
p'_b=\sum_{\mu\in\{0,x,y,z\}}\mathbb{E}_U\!\left[T_{b\mu}^2\right]p_\mu .
\label{eq:second_moment_reduced_derivation}
\end{equation}
Equivalently, following the four-component derivation explicitly, we first define
\begin{equation}
M_{\AD,b\mu}^{(4)}(c,\gamma)=\mathbb{E}_U\!\left[T_{\AD,b\mu}(U,\gamma)^2\right],
\end{equation}
with rows and columns ordered as $(0,x,y,z)$. The identity row gives $p'_0=p_0$. The nontrivial rows needed for the reduced variables are
\begin{align}
p'_z={}&\gamma^2 c\,p_0
+s\frac{1-c}{2}(p_x+p_y)+s^2c\,p_z,
\label{eq:AD_four_component_pz}\\
p'_x={}&\gamma^2\frac{1-c}{2}p_0
+s\frac{1+c}{4}(p_x+p_y)
+s^2\frac{1-c}{2}p_z,
\label{eq:AD_four_component_px}\\
p'_y={}&p'_x .
\label{eq:AD_four_component_py}
\end{align}
The reduction to the basis $(0,z,\perp)$ is made only at this point: we set $p_\perp=p_x+p_y$, use $p_x=p_y=p_\perp/2$, and take $p'_\perp=p'_x+p'_y$.

This projection yields

\begin{align}
p'_0&=p_0,\label{eq:AD_reduced_p0}\\
p'_z&=c\gamma^2p_0+cs^2p_z+\frac{1-c}{2}s\,p_\perp,\\
p'_\perp&=(1-c)\gamma^2p_0+(1-c)s^2p_z+\frac{1+c}{2}s\,p_\perp .
\label{eq:AD_reduced_moment_equations}
\end{align}

Equations~\eqref{eq:AD_reduced_p0}--\eqref{eq:AD_reduced_moment_equations} give the reduced transfer matrix. With $s=1-\gamma$, we write
\begin{equation}
R_{\AD}(c,\gamma)=
\begin{pmatrix}
1&0&0\\
c\gamma^2&cs^2&\frac{1-c}{2}s\\
(1-c)\gamma^2&(1-c)s^2&\frac{1+c}{2}s
\end{pmatrix}.
\label{eq:RAD}
\end{equation}
The first column is the nonunital source. It starts at order $\gamma^2$ in the local second-moment matrix, but it can affect stationary moments at lower apparent order because the radial gap itself is small at weak noise.

For the \DEP channel,
\begin{equation}
\Lambda_{\DEP,\gamma}(\omega)=(1-\gamma)\omega+\gamma\frac{\id}{2},
\end{equation}
so that the Bloch vector is multiplied by $\kappa=1-\gamma$. Hence
\begin{equation}
R_{\DEP}(c,\gamma)=
\begin{pmatrix}
1&0&0\\
0&\kappa^2c&\kappa^2\frac{1-c}{2}\\
0&\kappa^2(1-c)&\kappa^2\frac{1+c}{2}
\end{pmatrix}.
\label{eq:RDEP}
\end{equation}
For the \DPH channel in the $Z$ basis,
\begin{equation}
\Lambda_{\DPH,\gamma}(\omega)=(1-\gamma)\omega+\gamma\sigma_z\omega\sigma_z,
\end{equation}
which gives
\begin{equation}
c_x\mapsto \lambda c_x,\qquad
c_y\mapsto \lambda c_y,\qquad
c_z\mapsto c_z,
\qquad
\lambda=1-2\gamma.
\end{equation}
The reduced block is therefore
\begin{equation}
R_{\DPH}(c,\gamma)=
\begin{pmatrix}
1&0&0\\
0&c&\lambda^2\frac{1-c}{2}\\
0&1-c&\lambda^2\frac{1+c}{2}
\end{pmatrix}.
\label{eq:RDPH}
\end{equation}

\section{Radial and angular gaps}
\label{sec:radial_angular}

\subsection{Weak-noise criterion}

We use the same geometric terminology for the full noisy spectrum. The radial branch is the continuation of the isotropic non-identity mode, whose decay measures loss of Bloch radius or Pauli weight. The angular branch is the continuation of the anisotropy mode, whose decay measures redistribution among non-identity directions after the common radial contraction has been separated.

The transient block of Eq.~\eqref{eq:R0} is
\begin{equation}
B_0(c)=
\begin{pmatrix}
c&\frac{1-c}{2}\\
1-c&\frac{1+c}{2}
\end{pmatrix}.
\label{eq:B0}
\end{equation}
Its right eigenvectors may be chosen as
\begin{equation}
v_R=\begin{pmatrix}1\\2\end{pmatrix},
\qquad
v_A=\begin{pmatrix}1\\-1\end{pmatrix},
\end{equation}
with left eigenvectors
\begin{equation}
w_R^T=\frac{1}{3}(1,1),
\qquad
w_A^T=\frac{1}{3}(2,-1).
\end{equation}
Consider a general weak diagonal contraction before the rotation,
\begin{equation}
A(\gamma)=
\begin{pmatrix}
1-\alpha_z\gamma&0\\
0&1-\alpha_\perp\gamma
\end{pmatrix}
+O(\gamma^2),
\label{eq:general_contraction}
\end{equation}
so that $B(c,\gamma)=B_0(c)A(\gamma)$.

This equality denotes the composition acting on column vectors in the reduced $(p_z,p_\perp)$ sector: the diagonal noisy contraction acts first, and the averaged rotation block acts afterwards. To make the perturbative step explicit, we write
\[
A(\gamma)=I+\gamma A_1+O(\gamma^2),
\qquad
A_1=-\diag(\alpha_z,\alpha_\perp),
\]
and therefore
\[
B(c,\gamma)=B_0(c)+\gamma B_1(c)+O(\gamma^2),
\qquad
B_1(c)=B_0(c)A_1.
\]
For a simple eigenvalue of this generally non-Hermitian matrix, we expand
\[
\lambda_j(\gamma)=\lambda_j^{(0)}+\gamma\lambda_j^{(1)}+O(\gamma^2),
\qquad
j=R,A.
\]
With the normalization chosen above, $w_i^Tv_j=\delta_{ij}$, first-order perturbation theory gives
\[
\lambda_j^{(1)}=w_j^T B_1(c)v_j .
\]
Using $w_j^TB_0(c)=\lambda_j^{(0)}w_j^T$, one obtains
\[
\lambda_R^{(1)}
=w_R^TB_0(c)A_1v_R
=-\frac{\alpha_z+2\alpha_\perp}{3},
\]
and
\[
\lambda_A^{(1)}
=w_A^TB_0(c)A_1v_A
=-\mu_A(c)\frac{2\alpha_z+\alpha_\perp}{3}.
\]

Substituting these derivatives leads to
\begin{equation}
\lambda_R(\gamma)=1-\eta_R\gamma+O(\gamma^2),
\qquad
\eta_R=\frac{\alpha_z+2\alpha_\perp}{3},
\label{eq:etaR}
\end{equation}
and
\begin{equation}
\lambda_A(\gamma)=\mu_A(c)\left[1-\eta_A\gamma\right]+O(\gamma^2),
\qquad
\eta_A=\frac{2\alpha_z+\alpha_\perp}{3}.
\label{eq:etaA}
\end{equation}
Thus
\begin{equation}
\eta_A-\eta_R=\frac{\alpha_z-\alpha_\perp}{3}.
\label{eq:criterion}
\end{equation}
Then a local test for useful noise is: If $\alpha_z>\alpha_\perp$, the anisotropy of the channel suppresses the angular mode faster than the radial mode. If $\alpha_z=\alpha_\perp$, any increase of a raw gap is isotropic contraction. If $\alpha_z<\alpha_\perp$, the channel preferentially removes transverse components and the radial cost dominates the angular benefit.

The three channels fall into three different classes:
\begin{align}
\AD:\quad &\alpha_z=2,\quad \alpha_\perp=1,
&
\eta_R=\frac{4}{3},\quad \eta_A=\frac{5}{3},
\\
\DEP:\quad &\alpha_z=2,\quad \alpha_\perp=2,
&
\eta_R=2,\quad \eta_A=2,
\\
\DPH:\quad &\alpha_z=0,\quad \alpha_\perp=4,
&
\eta_R=\frac{8}{3},\quad \eta_A=\frac{4}{3}.
\end{align}
The \AD channel is therefore the only locally favorable channel among the three. The \DEP channel is neutral, and the \DPH channel is unfavorable in the sector-resolved sense.

\subsection{Universal radial leakage in the multicore circuit}

The radial mode is analytically tractable because, at zero noise, the uniform vector over all non-identity Pauli strings is stationary. Unitary gates, local or inter-core, only redistribute Pauli strings within the non-identity sector. Weak local noise is the only first-order mechanism that leaks weight out of that sector.

There are $4^N-1$ non-identity Pauli strings on $N$ qubits. For a fixed qubit, the number of such strings with $X$, $Y$, or $Z$ on that site is $4^{N-1}$ for each component. A one-qubit noisy operation with weak contraction coefficients $(\alpha_z,\alpha_\perp)$ therefore has the weighted average radial-loss coefficient

\[
\begin{aligned}
L_{\ch}(N)
&=
\frac{1}{4^N-1}
\Big[
0\cdot(4^{N-1}-1)
+\alpha_\perp 4^{N-1}
\\
&\hspace{3.0em}
+\alpha_\perp 4^{N-1}
+\alpha_z4^{N-1}
\Big].
\end{aligned}
\]
Here the average is taken over the radial ensemble, i.e., over all non-identity Pauli strings with equal weight. If the noisy qubit carries the identity, the string does not lose non-identity weight at first order; if it carries $X$ or $Y$, the loss coefficient is $\alpha_\perp$; if it carries $Z$, the loss coefficient is $\alpha_z$. The two identical $\alpha_\perp$ terms are the two transverse Pauli labels, $X$ and $Y$, equivalently the degeneracy compressed in $p_\perp=p_x+p_y$ and in the radial vector $(1,2)^T$. Thus $L_{\ch}(N)$ is not an additional dynamical assumption, but the Pauli-string average of the local contraction coefficient seen by the radial mode:

\begin{equation}
L_{\ch}(N)=
\frac{(2\alpha_\perp+\alpha_z)4^{N-1}}{4^N-1}.
\label{eq:Lch}
\end{equation}
Each local layer contains, on average, $p_1 C$ one-qubit noisy operations, and there are $k$ local layers per cycle. The inter-core gates are unitary and do not contribute to radial leakage at first order. Hence
\begin{equation}
\lambda_R^{\ch}(\gamma)
=
1-p_1Ck\,L_{\ch}(N)\gamma+O(\gamma^2).
\label{eq:lambdaR_cycle}
\end{equation}
The corresponding radial gap per elementary gate is
\begin{equation}
\frac{\Delta_R^{\ch}}{\gamma}
=
\frac{p_1}{1+r}
\frac{(2\alpha_\perp+\alpha_z)4^{N-1}}{4^N-1}
+O(\gamma).
\label{eq:universal_radial}
\end{equation}
For the three channels,
\begin{align}
\frac{\Delta_R^{\AD}}{\gamma}
&=
\frac{p_1}{1+r}\frac{4^N}{4^N-1}
+O(\gamma),
\\
\frac{\Delta_R^{\DEP}}{\gamma}
&=
\frac{p_1}{1+r}\frac{6\,4^{N-1}}{4^N-1}
+O(\gamma),
\\
\frac{\Delta_R^{\DPH}}{\gamma}
&=
\frac{p_1}{1+r}\frac{8\,4^{N-1}}{4^N-1}
+O(\gamma).
\end{align}
This hierarchy is important: among the three channels, \AD has the smallest radial leakage coefficient, \DEP is intermediate, and \DPH has the largest radial contribution.

\subsection{Diagnostics for noisy Haar mixing}

The angular sector cannot be reduced to a universal counting argument. Its dominant branch depends on how the local randomization, the intra-core gates, and the inter-core graph redistribute Pauli strings. We therefore track the relevant branch numerically.

For each point of parameter space we define
\begin{equation}
\Delta_A=1-|\lambda_A|^{1/D},
\label{eq:angular_gap}
\end{equation}
where $\lambda_A$ is the continued angular branch, and
\begin{equation}
\Delta_R=1-\lambda_R^{1/D}.
\label{eq:radial_gap}
\end{equation}
The ratio
\begin{equation}
\rho=\frac{\Delta_R}{\Delta_A}
\label{eq:dissipative_ratio}
\end{equation}
measures how much radial decay accompanies one unit of angular relaxation. We use $\rho<0.1$ as a weakly dissipative regime, $0.1\lesssim\rho\lesssim0.3$ as an intermediate regime, and $\rho\gtrsim0.3$ as a regime in which radial decay is no longer a small correction.

We also use a more conservative radial-normalized angular gap,
\begin{equation}
\Delta_{A|R}
=
1-\left(\frac{|\lambda_A|}{\lambda_R}\right)^{1/D}.
\label{eq:relative_angular_gap}
\end{equation}
This quantity factors out the common radial contraction. If a channel merely shrinks all non-identity moments, a raw increase in $\Delta_A$ disappears or reverses in $\Delta_{A|R}$.
In the numerical summaries below, a superscript $*$ denotes optimization over the scanned $(c,k)$ grid at fixed channel, topology, partition, and noise strength. For $\rho^*$, the value is evaluated at the same $(c,k)$ point that maximizes $\Delta_A$, so that it measures the radial cost of the angular optimum.

\section{Numerical results}
\label{sec:results}

\subsection{Branch-resolved data set}

The numerical construction follows the noiseless multicore Markov framework of Ref.~\cite{MontesPRA2026}. The local block $R_0(c)$ is replaced by $R_{\ch}(c,\gamma)$, and the full transfer matrix is constructed in the reduced $3^N$ basis. The refined scans used in the figures take
\begin{equation}
p_1=\frac{1}{2},\qquad
c\in\{0.1,0.2,\ldots,0.9,1/3,0.95\},
\end{equation}
\begin{equation}
\gamma\in[0,2\times10^{-2}],
\qquad
k=1,\ldots,14.
\end{equation}
The noise mesh is denser close to zero and includes $\gamma=10^{-4}$, $2\times10^{-4}$, $5\times10^{-4}$, and $10^{-3}$. We analyze the linear, star, ring, and fully connected topologies for the available branch-resolved partitions $2\times2$, $2\times3$, and $3\times2$. This gives $36$ refined tensors: three channels, four topologies, and three partitions.

The gate count used in Eq.~\eqref{eq:gapdef} matches the numerical implementation: $E_G=C-1$ for the linear and star graphs, $E_G=C$ for the ring, and $E_G=C(C-1)/2$ for the fully connected graph. Thus, for $C=2$, the ring case is treated as a denser communication layer on the same pair of cores rather than as a single linear link.

Branch identification is crucial. At $\gamma=0$ there is a cluster of unit eigenvalues associated with the identity and with the non-identity radial mode. For $\gamma>0$, the radial eigenvalue separates as $\lambda_R=1-O(\gamma)$. The angular branch is initialized at $\gamma=0$ as the dominant nonstationary eigenvalue outside the unit cluster and is then continued in $\gamma$ by proximity in the complex plane. In all refined data, the maximum branch jump is smaller than $4.8\times10^{-2}$, well below the numerical guard value $0.25$.

\begin{figure*}[t]
\centering
\includegraphics[width=0.90\textwidth]{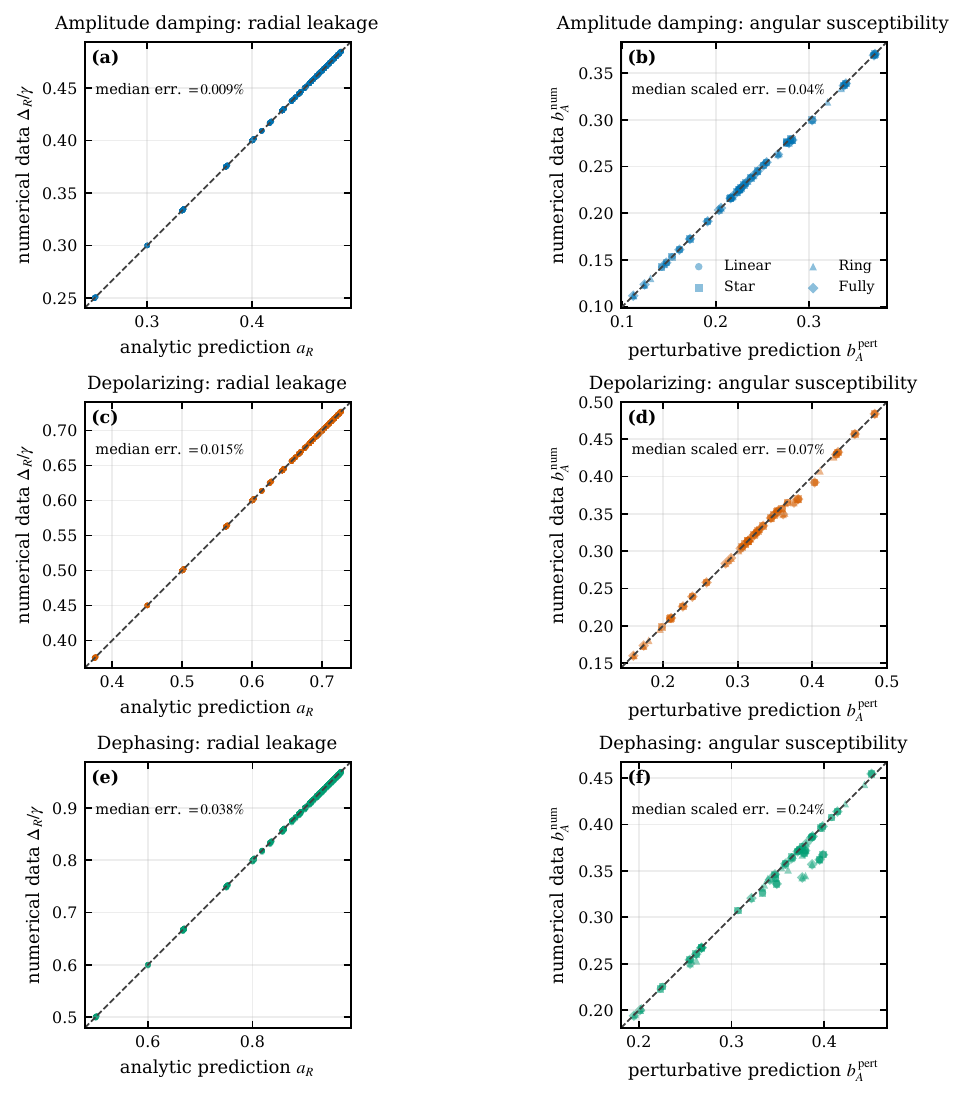}
\caption{Validation of the analytical predictions in the weak-noise regime. The left column tests the universal radial law, Eq.~\eqref{eq:universal_radial}, by comparing the numerical coefficient $\Delta_R/\gamma$ with the analytical prediction $a_R$ at $\gamma=10^{-4}$ for the available topology, partition, $c$, and $k$ grid. The right column tests the branch-resolved angular susceptibility by comparing $b_A^{\rm num}=[\Delta_A(10^{-4})-\Delta_A(0)]/10^{-4}$ with the perturbative prediction $b_A^{\rm pert}$ obtained from the full derivative $M'_{\ch}(0)$ on a sample of $324$ points. Panels (a,c,e) correspond to the radial test for \AD, \DEP, and \DPH, respectively; panels (b,d,f) show the corresponding angular-susceptibility test. Dashed lines denote perfect agreement.}
\label{fig:analytic_validation}
\end{figure*}

Figure~\ref{fig:analytic_validation} validates the two analytical ingredients used in the rest of the paper. In the radial sector, the data collapse on the diagonal because the coefficient depends only on the channel, $N$, and $r=E_G/(Ck)$, and not on the local rotation parameter $c$. At $\gamma=10^{-4}$, the median relative errors are $0.009\%$, $0.015\%$, and $0.038\%$ for \AD, \DEP, and \DPH. In the angular sector there is no analogous closed counting law, so the analytical object we test is the weak-noise slope of the angular gap for each fixed architecture point.

For fixed topology, partition, $c$, and $k$, the full branch-resolved transfer matrix is expanded as
\[
M_{\ch}(\gamma)=M_0+\gamma M_{\ch}^{(1)}+O(\gamma^2),
\qquad
M_{\ch}^{(1)}=M'_{\ch}(0).
\]
The derivative $M_{\ch}^{(1)}$ is obtained analytically by differentiating the one-qubit noisy blocks and propagating that derivative through the same matrix products used to build $M_{\ch}(\gamma)$. We then choose the angular eigenvalue $\lambda_{A,0}$ of $M_0$ as the dominant nonstationary branch outside the radial unit cluster and linearize only that branch,
\[
\lambda_A(\gamma)=\lambda_{A,0}+\gamma\lambda_A'(0)+O(\gamma^2).
\]
This is the angular approximation used in the right column of Fig.~\ref{fig:analytic_validation}: it is a local-in-$\gamma$ susceptibility, not a closed universal formula for the angular branch at finite noise.

If $v_A$ and $w_A$ are the right and left eigenvectors of this angular branch at $\gamma=0$, then
\begin{equation}
\lambda_A'(0)=
\frac{w_A^\dagger M'_{\ch}(0)v_A}{w_A^\dagger v_A}.
\end{equation}

The quantity plotted is obtained after converting this eigenvalue derivative into the slope of the per-gate angular gap,
\[
\Delta_A(\gamma)
=
1-|\lambda_A(\gamma)|^{1/D}
=
\Delta_A(0)+b_A^{\rm pert}\gamma+O(\gamma^2),
\]
with
\[
b_A^{\rm pert}
=
-\frac{|\lambda_{A,0}|^{1/D}}{D}
\operatorname{Re}\!\left[\frac{\lambda_A'(0)}{\lambda_{A,0}}\right],
\]
for the nonzero angular eigenvalues used in the sampled data.

With this definition of $b_A^{\rm pert}$, the median scaled errors in the angular susceptibility are $0.041\%$, $0.073\%$, and $0.235\%$ for \AD, \DEP, and \DPH. This confirms both the local noise matrices and the branch continuation procedure.

\subsection{Raw angular speed-up versus dissipative cost}

\begin{figure*}[t]
\centering
\includegraphics[width=0.98\textwidth]{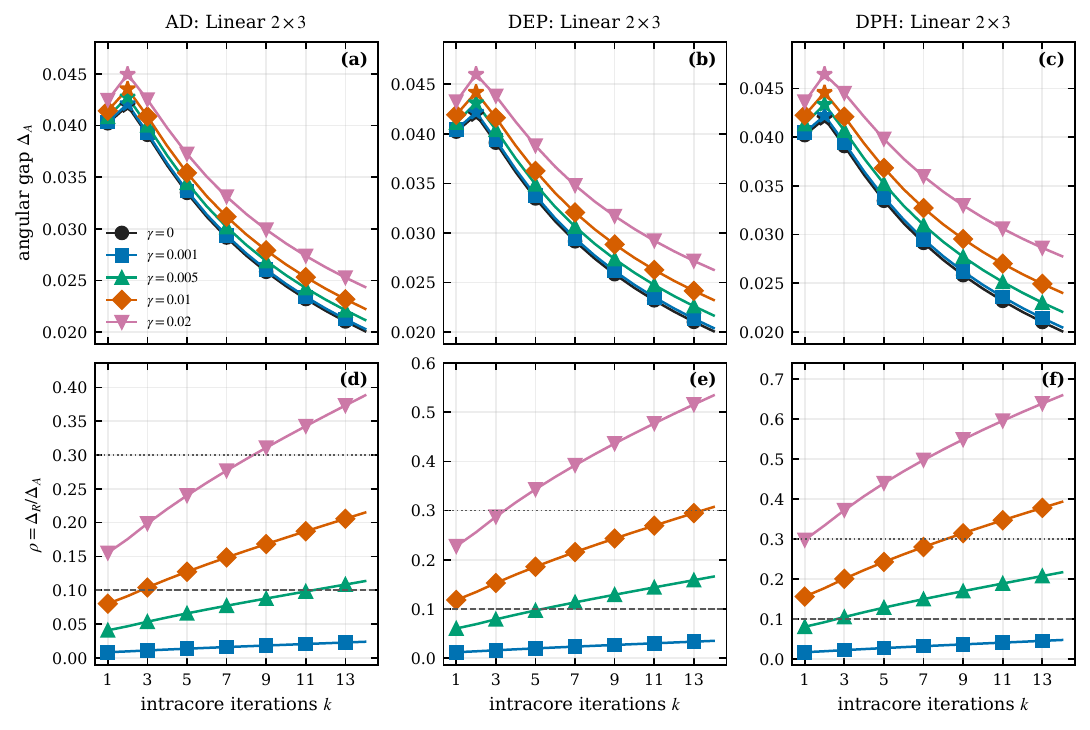}
\caption{Representative angular gap curves as a function of the number of intra-core iterations for a linear $2\times3$ architecture. For each $k$ the gap is optimized over the grid of $c$. The upper row shows the branch-resolved angular gap $\Delta_A$ for the three channels. The lower row shows the dissipative ratio $\rho=\Delta_R/\Delta_A$ for the same configurations. Horizontal lines mark $\rho=0.1$ and $\rho=0.3$.}
\label{fig:refined_gap_ipc_channels}
\end{figure*}

Figure~\ref{fig:refined_gap_ipc_channels} illustrates why the noisy spectrum must be resolved into sectors. Looking only at the angular branch, all three channels increase $\Delta_A$ as $\gamma$ grows. This does not imply that all three channels improve Haar mixing in the same sense. The radial cost, quantified by the dissipative ratio $\rho$, follows the robust ordering
\begin{equation}
\rho_{\AD}<\rho_{\DEP}<\rho_{\DPH}.
\end{equation}
\DEP and \DPH can therefore look favorable in a raw gap plot while paying a larger radial cost. In the linear $2\times3$ example, at $\gamma=0.020$ \DPH exceeds the dissipative threshold $\rho=0.3$ over a broad part of the curve, whereas \AD remains closer to the intermediate regime.

\begin{figure*}[t]
\centering
\includegraphics[width=0.98\textwidth]{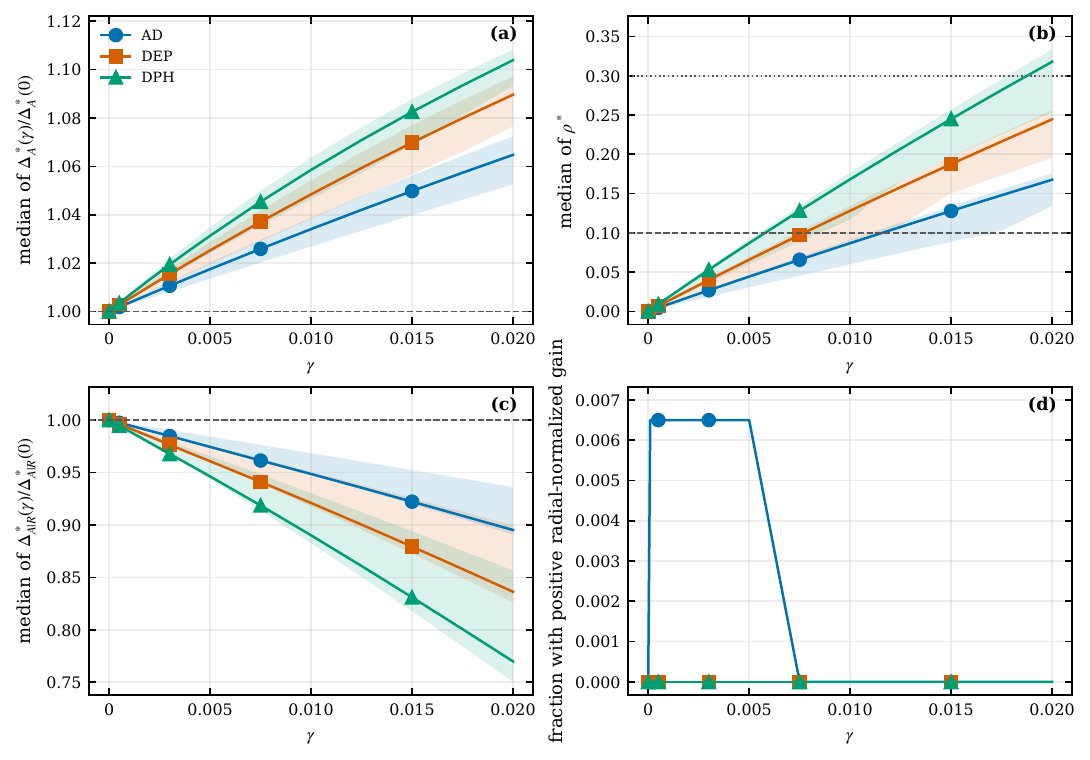}
\caption{Statistical summary over the $36$ refined tensors. (a) Median and interquartile range of the optimized angular improvement $\Delta_A^*(\gamma)/\Delta_A^*(0)$. (b) Median dissipative ratio $\rho^*$ at the angular optimum. (c) Radial-normalized angular gap $\Delta_{A|R}^*(\gamma)/\Delta_{A|R}^*(0)$. (d) Fraction of $(c,k)$ grid points with positive radial-normalized improvement. The superscript $*$ denotes optimization over the scanned $(c,k)$ grid, except for $\rho^*$, which is evaluated at the angular-gap optimum.}
\label{fig:refined_channel_summary}
\end{figure*}

Figure~\ref{fig:refined_channel_summary} summarizes the same point over the full data set. Panel (a) shows that the optimized raw angular gap increases for all channels. At $\gamma=10^{-3}$, the median gains are approximately $0.36\%$ for \AD, $0.52\%$ for \DEP, and $0.67\%$ for \DPH. At $\gamma=0.020$, they reach $6.5\%$, $9.0\%$, and $10.4\%$. A raw-gap analysis alone would therefore identify \DPH as the most beneficial channel.

Panel (b) changes the interpretation. At $\gamma=10^{-3}$, the median dissipative ratios are $0.0090$, $0.0135$, and $0.0179$ for \AD, \DEP, and \DPH. At $\gamma=0.020$, they grow to $0.168$, $0.245$, and $0.318$. \AD produces a smaller raw angular gain, but it does so with the smallest radial cost. \DPH produces the largest raw angular gain, but enters the dissipative regime first.

Panels (c) and (d) provide the most conservative test. Once the radial mode is factored out through Eq.~\eqref{eq:relative_angular_gap}, the median of $\Delta_{A|R}^*/\Delta_{A|R}^*(0)$ decreases for all channels. This shows that a significant part of the raw gain in $\Delta_A$ is common radial contraction. Nevertheless, panel (d) reveals a qualitative difference: \AD has the clearest positive weak-noise signal, although only on a small fraction of the parameter grid. These points are concentrated near $c=0.95$ and $k=1$, where the noiseless local layer is deliberately slow to mix. \DEP shows no robust positive signal in the statistical summary, and \DPH shows no positive radial-normalized improvement in the refined mesh. This supports the analytical mechanism without overstating it: \AD opens the most consistent narrow window of genuine angular improvement, but the useful effect is limited by radial leakage.

\subsection{Noise-renormalized communication optimum}

\begin{figure*}[t]
\centering
\includegraphics[width=0.98\textwidth]{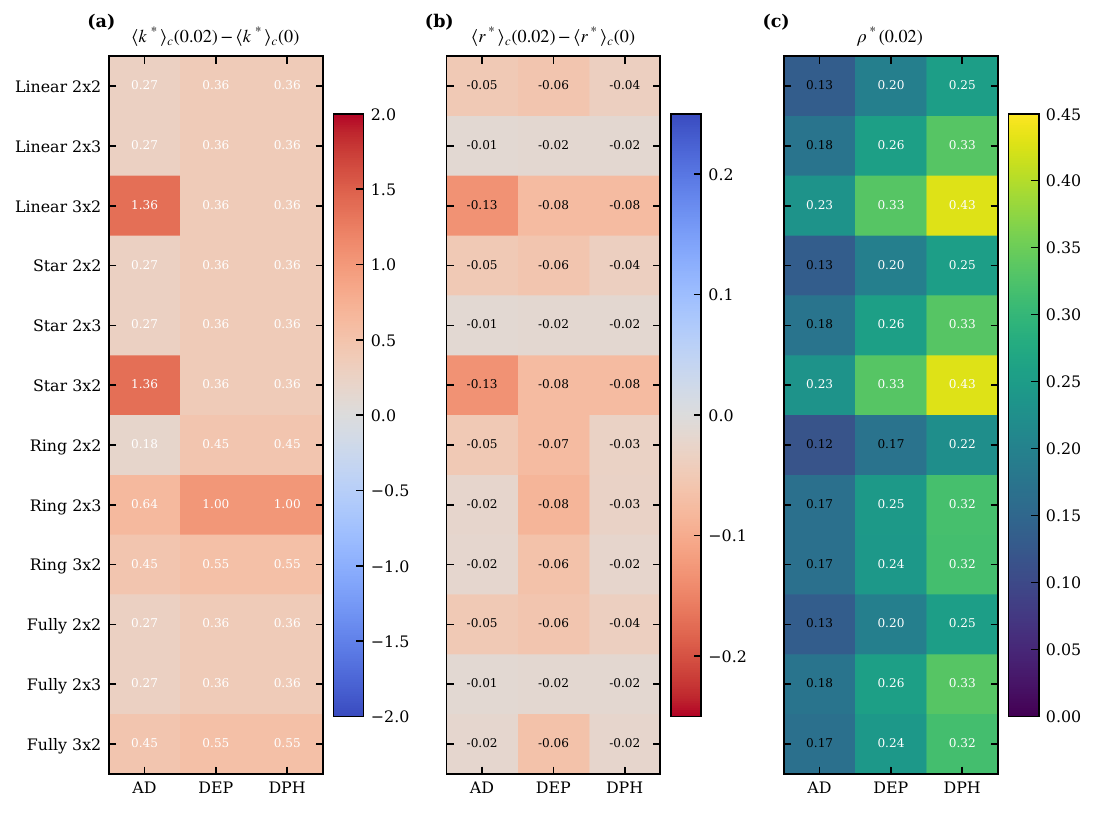}
\caption{Change in the optimal configuration between $\gamma=0$ and $\gamma=0.020$. For each architecture and channel, the local depth $k$ is optimized for each value of $c$ and then averaged over $c$. (a) Shift of the mean optimal number of intra-core iterations. (b) Change in the inter-core communication ratio $r=E_G/(Ck)$. (c) Dissipative ratio at the global angular optimum for $\gamma=0.020$.}
\label{fig:refined_configuration_heatmaps}
\end{figure*}

Figure~\ref{fig:refined_configuration_heatmaps} shows that noise also renormalizes the optimal architecture parameters. As $\gamma$ increases to $0.020$, the conditional optimum $k^*(c)$ generally shifts toward more intra-core iterations. The shift is topology and partition dependent. For \AD, the linear/star $3\times2$ architectures increase the average $\langle k^*\rangle_c$ by approximately $1.36$, whereas the $2\times2$ architectures change by only about $0.27$. In terms of Eq.~\eqref{eq:r_definition}, this corresponds to a reduction of the average communication ratio: as radial cost grows, the angular optimum tends to communicate slightly less frequently per local gate. Panel (c) again displays the channel hierarchy. At $\gamma=0.020$, \AD remains below $\rho\simeq0.24$ in all cases, \DEP reaches values up to $\rho\simeq0.33$, and \DPH reaches values up to $\rho\simeq0.43$ in the most sensitive configurations.

The conclusion from the numerical study is therefore sectorial. Noise can increase the branch-resolved angular gap, but a raw increase of $\Delta_A$ is not by itself evidence of improved approach to Haar statistics. The favorable case is the one in which the angular gain remains visible after the common radial contraction is removed. Among the three channels studied here, \AD is the closest to that regime.

\section{Discussion and conclusions}
\label{sec:discussion_conclusions}

The main conceptual conclusion of our work is that noisy Haar mixing must be resolved by branch. In a unitary random circuit, the Markov gap measures convergence inside the conserved non-identity Pauli sector. In an open circuit, the same term ``gap'' can describe two different effects: angular randomization within the non-identity sector, or radial leakage out of it. Treating these processes as one can lead to the misleading conclusion that any noise helps simply because it reduces eigenvalue moduli.

\AD illustrates the nontrivial possibility. It is dissipative and nonunital, and therefore it certainly damages purity. However, its homogeneous part contracts longitudinal second moments more strongly than transverse ones. Combined with random rotations, this produces a weak-noise angular advantage according to Eq.~\eqref{eq:criterion}. The numerical data show that this advantage is not global or unconditional: the radial-normalized improvement appears only in a small region of the parameter grid. Still, \AD has the most consistent positive signal and the smallest radial cost at the angular optima.

\DEP illustrates the trivial case. It contracts all Bloch directions equally, so it can increase a raw angular gap without producing a genuine angular advantage once the radial mode is factored out.

\DPH is anisotropic, but with the wrong sign for this problem. It suppresses transverse components, produces the largest raw angular gap increase, and also produces the largest dissipative ratio.

For distributed architectures, the split between radial and angular sectors is especially useful. The radial leakage coefficient is universal at first order and depends only on the channel, the number of qubits, and the communication ratio. The angular optimum remains architecture dependent and must be computed or bounded for each topology and partition. Thus the channel fixes the dissipative budget, while the architecture determines whether that budget can be converted into useful angular mixing before radial decay dominates.

In summary, we have extended the second-moment transfer-matrix theory of distributed random circuits to open dynamics with \AD, \DEP, and \DPH. We derived reduced one-qubit transfer matrices, identified radial and angular spectral branches, obtained a universal weak-noise radial law by counting Pauli strings, and introduced a radial-normalized angular diagnostic. The numerical analysis of $36$ branch-resolved tensors shows that raw angular gaps increase for all three channels, but that genuine angular improvement is confined to narrow weak-noise windows, most consistently for \AD. The result is a cautious form of noise assistance: weak noise can help the angular part of Haar mixing only when its anisotropy is favorable and the distributed architecture uses that anisotropy before radial dissipation becomes the dominant process.

\begin{acknowledgments}
This work was partially supported by the Ministerio de Ciencia, Innovación y Universidades, Gobierno de España, under contract No. PID2021-122711NB-C21. We also acknowledge support from CONICET.
\end{acknowledgments}

\section*{Data availability}

The numerical data and analysis scripts supporting the figures will be made available in a public repository associated with the final version of the manuscript. Until then, they are available from the authors upon reasonable request. The refined tensors include the three noise channels and the four network topologies, together with the launcher that regenerates the figures from the stored tensors.

\appendix

\section{Complete one-qubit derivation of the noisy reduced transfer blocks}
\label{app:complete_noise_blocks}

\subsection{Common second-moment construction}
\label{app:common_second_moment_construction}

We derive the reduced one-qubit blocks in the Schrödinger picture. The one-qubit state is written as
\begin{equation}
\omega=\frac{1}{2}\left(\sigma_0+\sum_{a=x,y,z}c_a\sigma_a\right),
\qquad
c_a=\Tr(\sigma_a\omega),
\end{equation}
and we keep the identity coefficient
\begin{equation}
c_0=\Tr(\sigma_0\omega)=1,\qquad
\tilde{\bm c}=(c_0,c_x,c_y,c_z)^T .
\end{equation}
The four second moments are
\begin{equation}
p_\mu=\mathbb{E}(c_\mu^2),\qquad \mu\in\{0,x,y,z\}.
\end{equation}
As in the reduced second-moment closure used in the main text, we assume no cross correlations,
\begin{equation}
\mathbb{E}(c_\mu c_\nu)=0,\qquad \mu\neq\nu,
\end{equation}
and transverse symmetry,
\begin{equation}
p_x=p_y,\qquad p_\perp=p_x+p_y .
\end{equation}

In one elementary noisy local step, the channel acts before the random unitary,
\begin{equation}
\omega'=U\,\Lambda_{\ch,\gamma}(\omega)\,U^\dagger .
\end{equation}
The unitary induces a real rotation $O(U)\in SO(3)$ on the non-identity Bloch components and leaves the identity invariant,
\begin{equation}
L(U)=
\begin{pmatrix}
1&0&0&0\\
0&O_{xx}&O_{xy}&O_{xz}\\
0&O_{yx}&O_{yy}&O_{yz}\\
0&O_{zx}&O_{zy}&O_{zz}
\end{pmatrix}.
\end{equation}
The ensemble of rotations is described by
\begin{align}
\mathbb{E}(O_{zz}^2)&=c,\\
\mathbb{E}(O_{xz}^2)=\mathbb{E}(O_{yz}^2)&=\frac{1-c}{2},\\
\mathbb{E}(O_{zx}^2)=\mathbb{E}(O_{zy}^2)&=\frac{1-c}{2},\\
\mathbb{E}(O_{xx}^2)=\mathbb{E}(O_{yy}^2)
=\mathbb{E}(O_{xy}^2)=\mathbb{E}(O_{yx}^2)&=\frac{1+c}{4}.
\end{align}

For each noise channel, we first write its four-component Bloch map as
\begin{equation}
\tilde{\bm c}^{\,\ch}=A_{\ch}^{(4)}(\gamma)\tilde{\bm c}.
\end{equation}
After the rotation,
\begin{equation}
\tilde{\bm c}'=T_{\ch}(U,\gamma)\tilde{\bm c},
\qquad
T_{\ch}(U,\gamma)=L(U)A_{\ch}^{(4)}(\gamma),
\end{equation}
or componentwise
\begin{equation}
c'_b=\sum_{\mu\in\{0,x,y,z\}}T_{\ch,b\mu}(U,\gamma)c_\mu .
\end{equation}
The four-component transfer matrix for second moments is then
\begin{equation}
M_{\ch,b\mu}^{(4)}(c,\gamma)=
\mathbb{E}_U\!\left[T_{\ch,b\mu}(U,\gamma)^2\right],
\end{equation}
so that
\begin{equation}
p'_b=\sum_{\mu\in\{0,x,y,z\}}M_{\ch,b\mu}^{(4)}(c,\gamma)p_\mu .
\label{eq:app_four_component_moment_rule}
\end{equation}
Only after this step do we reduce to $(0,z,\perp)$ by setting
$p_x=p_y=p_\perp/2$ and $p'_\perp=p'_x+p'_y$.

\subsection{Amplitude damping}
\label{app:complete_ad_derivation}

For amplitude damping,
\begin{equation}
K_0=
\begin{pmatrix}
1&0\\
0&\sqrt{1-\gamma}
\end{pmatrix},
\qquad
K_1=
\begin{pmatrix}
0&\sqrt{\gamma}\\
0&0
\end{pmatrix},
\end{equation}
and
\begin{equation}
\Lambda_{\AD,\gamma}(\omega)
=K_0\omega K_0^\dagger+K_1\omega K_1^\dagger .
\end{equation}
Writing $s=1-\gamma$, the corresponding four-component Bloch map is
\begin{equation}
\begin{pmatrix}
c_0\\ c_x\\ c_y\\ c_z
\end{pmatrix}
\mapsto
\begin{pmatrix}
c_0\\
\sqrt{s}\,c_x\\
\sqrt{s}\,c_y\\
s\,c_z+\gamma c_0
\end{pmatrix},
\end{equation}
or
\begin{equation}
A_{\AD}^{(4)}(\gamma)=
\begin{pmatrix}
1&0&0&0\\
0&\sqrt{s}&0&0\\
0&0&\sqrt{s}&0\\
\gamma&0&0&s
\end{pmatrix}.
\end{equation}
The composed transfer matrix is
\begin{equation}
T_{\AD}(U,\gamma)=
\begin{pmatrix}
1&0&0&0\\
\gamma O_{xz}&\sqrt{s}O_{xx}&\sqrt{s}O_{xy}&sO_{xz}\\
\gamma O_{yz}&\sqrt{s}O_{yx}&\sqrt{s}O_{yy}&sO_{yz}\\
\gamma O_{zz}&\sqrt{s}O_{zx}&\sqrt{s}O_{zy}&sO_{zz}
\end{pmatrix}.
\end{equation}
Squaring each entry and using the rotation averages gives
\begin{equation}
M_{\AD}^{(4)}(c,\gamma)=
\begin{pmatrix}
1&0&0&0\\
\gamma^2\tfrac{1-c}{2}&s\tfrac{1+c}{4}&s\tfrac{1+c}{4}&s^2\tfrac{1-c}{2}\\
\gamma^2\tfrac{1-c}{2}&s\tfrac{1+c}{4}&s\tfrac{1+c}{4}&s^2\tfrac{1-c}{2}\\
\gamma^2c&s\tfrac{1-c}{2}&s\tfrac{1-c}{2}&s^2c
\end{pmatrix},
\label{eq:app_MAD_four}
\end{equation}
with rows and columns ordered as $(0,x,y,z)$. Equivalently, the four-component moment equations needed for the reduction are
\begin{align}
p'_0&=p_0,\\
p'_z&=\gamma^2c\,p_0+s\tfrac{1-c}{2}(p_x+p_y)+s^2c\,p_z,\\
p'_x&=\gamma^2\tfrac{1-c}{2}p_0
+s\tfrac{1+c}{4}(p_x+p_y)
+s^2\tfrac{1-c}{2}p_z,\\
p'_y&=p'_x .
\end{align}
Using $p_\perp=p_x+p_y$ and $p'_\perp=p'_x+p'_y$, one obtains
\begin{align}
p'_0&=p_0,\\
p'_z&=c\gamma^2p_0+cs^2p_z+\tfrac{1-c}{2}s\,p_\perp,\\
p'_\perp&=(1-c)\gamma^2p_0+(1-c)s^2p_z+\tfrac{1+c}{2}s\,p_\perp .
\end{align}
Therefore,
\begin{equation}
R_{\AD}(c,\gamma)=
\begin{pmatrix}
1&0&0\\
c\gamma^2&cs^2&\tfrac{1-c}{2}s\\
(1-c)\gamma^2&(1-c)s^2&\tfrac{1+c}{2}s
\end{pmatrix}.
\label{eq:app_RAD}
\end{equation}
For $\gamma=0$, this reduces to the noiseless block $R_0(c)$.

\subsection{Depolarizing noise}
\label{app:complete_dep_derivation}

For depolarizing noise,
\begin{equation}
\Lambda_{\DEP,\gamma}(\omega)
=(1-\gamma)\omega+\gamma\frac{\id}{2},
\qquad
\kappa=1-\gamma .
\end{equation}
In the Pauli expansion this gives
\begin{equation}
\begin{pmatrix}
c_0\\ c_x\\ c_y\\ c_z
\end{pmatrix}
\mapsto
\begin{pmatrix}
c_0\\
\kappa c_x\\
\kappa c_y\\
\kappa c_z
\end{pmatrix},
\end{equation}
or
\begin{equation}
A_{\DEP}^{(4)}(\gamma)=
\begin{pmatrix}
1&0&0&0\\
0&\kappa&0&0\\
0&0&\kappa&0\\
0&0&0&\kappa
\end{pmatrix}.
\end{equation}
The composed transfer matrix is
\begin{equation}
T_{\DEP}(U,\gamma)=
\begin{pmatrix}
1&0&0&0\\
0&\kappa O_{xx}&\kappa O_{xy}&\kappa O_{xz}\\
0&\kappa O_{yx}&\kappa O_{yy}&\kappa O_{yz}\\
0&\kappa O_{zx}&\kappa O_{zy}&\kappa O_{zz}
\end{pmatrix}.
\end{equation}
Squaring and averaging gives
\begin{equation}
M_{\DEP}^{(4)}(c,\gamma)=
\begin{pmatrix}
1&0&0&0\\
0&\kappa^2\tfrac{1+c}{4}&\kappa^2\tfrac{1+c}{4}&\kappa^2\tfrac{1-c}{2}\\
0&\kappa^2\tfrac{1+c}{4}&\kappa^2\tfrac{1+c}{4}&\kappa^2\tfrac{1-c}{2}\\
0&\kappa^2\tfrac{1-c}{2}&\kappa^2\tfrac{1-c}{2}&\kappa^2c
\end{pmatrix}.
\label{eq:app_MDEP_four}
\end{equation}
The four-component moment equations are
\begin{align}
p'_0&=p_0,\\
p'_z&=\kappa^2\!\left[c\,p_z+\tfrac{1-c}{2}(p_x+p_y)\right],\\
p'_x&=\kappa^2\!\left[\tfrac{1+c}{4}(p_x+p_y)+\tfrac{1-c}{2}p_z\right],\\
p'_y&=p'_x .
\end{align}
Therefore,
\begin{align}
p'_0&=p_0,\\
p'_z&=\kappa^2\!\left(c\,p_z+\tfrac{1-c}{2}p_\perp\right),\\
p'_\perp&=\kappa^2\!\left((1-c)p_z+\tfrac{1+c}{2}p_\perp\right),
\end{align}
and the reduced block is
\begin{equation}
R_{\DEP}(c,\gamma)=
\begin{pmatrix}
1&0&0\\
0&\kappa^2c&\kappa^2\tfrac{1-c}{2}\\
0&\kappa^2(1-c)&\kappa^2\tfrac{1+c}{2}
\end{pmatrix}.
\label{eq:app_RDEP}
\end{equation}
Equivalently, the channel first multiplies the non-identity second moments by $\kappa^2$ and the rotation then applies the noiseless mixing block $R_0(c)$.

\subsection{Dephasing noise}
\label{app:complete_dph_derivation}

For dephasing in the $Z$ basis,
\begin{equation}
\Lambda_{\DPH,\gamma}(\omega)
=(1-\gamma)\omega+\gamma\sigma_z\omega\sigma_z,
\qquad
\lambda=1-2\gamma .
\end{equation}
Using $\sigma_z\sigma_x\sigma_z=-\sigma_x$,
$\sigma_z\sigma_y\sigma_z=-\sigma_y$, and
$\sigma_z\sigma_z\sigma_z=\sigma_z$, the Bloch components transform as
\begin{equation}
\begin{pmatrix}
c_0\\ c_x\\ c_y\\ c_z
\end{pmatrix}
\mapsto
\begin{pmatrix}
c_0\\
\lambda c_x\\
\lambda c_y\\
c_z
\end{pmatrix}.
\end{equation}
Thus
\begin{equation}
A_{\DPH}^{(4)}(\gamma)=
\begin{pmatrix}
1&0&0&0\\
0&\lambda&0&0\\
0&0&\lambda&0\\
0&0&0&1
\end{pmatrix}.
\end{equation}
The composed transfer matrix is
\begin{equation}
T_{\DPH}(U,\gamma)=
\begin{pmatrix}
1&0&0&0\\
0&\lambda O_{xx}&\lambda O_{xy}&O_{xz}\\
0&\lambda O_{yx}&\lambda O_{yy}&O_{yz}\\
0&\lambda O_{zx}&\lambda O_{zy}&O_{zz}
\end{pmatrix}.
\end{equation}
The averaged second-moment matrix is
\begin{equation}
M_{\DPH}^{(4)}(c,\gamma)=
\begin{pmatrix}
1&0&0&0\\
0&\lambda^2\tfrac{1+c}{4}&\lambda^2\tfrac{1+c}{4}&\tfrac{1-c}{2}\\
0&\lambda^2\tfrac{1+c}{4}&\lambda^2\tfrac{1+c}{4}&\tfrac{1-c}{2}\\
0&\lambda^2\tfrac{1-c}{2}&\lambda^2\tfrac{1-c}{2}&c
\end{pmatrix}.
\label{eq:app_MDPH_four}
\end{equation}
Therefore,
\begin{align}
p'_0&=p_0,\\
p'_z&=c\,p_z+\lambda^2\tfrac{1-c}{2}(p_x+p_y),\\
p'_x&=\lambda^2\tfrac{1+c}{4}(p_x+p_y)+\tfrac{1-c}{2}p_z,\\
p'_y&=p'_x .
\end{align}
Reducing to $(0,z,\perp)$ gives
\begin{align}
p'_0&=p_0,\\
p'_z&=c\,p_z+\lambda^2\tfrac{1-c}{2}p_\perp,\\
p'_\perp&=(1-c)p_z+\lambda^2\tfrac{1+c}{2}p_\perp,
\end{align}
and hence
\begin{equation}
R_{\DPH}(c,\gamma)=
\begin{pmatrix}
1&0&0\\
0&c&\lambda^2\tfrac{1-c}{2}\\
0&1-c&\lambda^2\tfrac{1+c}{2}
\end{pmatrix}.
\label{eq:app_RDPH}
\end{equation}
This completes the parallel derivation of the three noisy one-qubit transfer blocks used in the main text.

\bibliographystyle{apsrev4-2}
\bibliography{references}

@article{GoogleQuantumAI2025,
  author = {{Google Quantum AI and Collaborators}},
  title = {Quantum error correction below the surface code threshold},
  journal = {Nature},
  volume = {638},
  pages = {920},
  year = {2025}
}

@article{Bravyi2024,
  author = {Bravyi, S. and Cross, A. and Gambetta, J. and Maslov, D. and Rall, P. and Yoder, T.},
  title = {High-threshold and low-overhead fault-tolerant quantum memory},
  journal = {Nature},
  volume = {627},
  pages = {778},
  year = {2024}
}

@article{Bland2025,
  author = {Bland, M. and others},
  title = {Fault-tolerant quantum computing with improved quantum codes},
  journal = {Nature},
  volume = {647},
  pages = {343},
  year = {2025}
}

@article{Preskill2018,
  author = {Preskill, John},
  title = {Quantum Computing in the {NISQ} era and beyond},
  journal = {Quantum},
  volume = {2},
  pages = {79},
  year = {2018}
}

@article{Arute2019,
  author = {Arute, F. and others},
  title = {Quantum supremacy using a programmable superconducting processor},
  journal = {Nature},
  volume = {574},
  pages = {505},
  year = {2019}
}

@article{Pan2022,
  author = {Pan, Feng and Zhang, Pan},
  title = {Simulating the {Sycamore} quantum supremacy circuits},
  journal = {Phys. Rev. Lett.},
  volume = {128},
  pages = {030501},
  year = {2022}
}

@article{King2025,
  author = {King, A. D. and others},
  title = {Computational quantum advantage with quantum annealing},
  journal = {Science},
  volume = {388},
  pages = {199},
  year = {2025}
}

@article{Madsen2022,
  author = {Madsen, L. S. and others},
  title = {Quantum computational advantage with a programmable photonic processor},
  journal = {Nature},
  volume = {606},
  pages = {75},
  year = {2022}
}

@article{Bluvstein2024,
  author = {Bluvstein, D. and others},
  title = {Logical quantum processor based on reconfigurable atom arrays},
  journal = {Nature},
  volume = {626},
  pages = {58},
  year = {2024}
}

@article{Rad2025,
  author = {Rad, H. A. and others},
  title = {A scalable photonic quantum computer},
  journal = {Nature},
  volume = {638},
  pages = {912},
  year = {2025}
}

@article{Jnane2022,
  author = {Jnane, H. and Undseth, B. and Cai, Z. and Benjamin, S. C. and Koczor, B.},
  title = {Multicore quantum computing},
  journal = {Phys. Rev. Applied},
  volume = {18},
  pages = {044064},
  year = {2022}
}

@article{Hetenyi2024,
  author = {Het{\'e}nyi, B. and Wootton, J. R.},
  title = {Tailoring quantum circuits to distributed architectures},
  journal = {PRX Quantum},
  volume = {5},
  pages = {040334},
  year = {2024}
}

@article{Wu2024,
  author = {Wu, X. and others},
  title = {A modular quantum processor with a reconfigurable router},
  journal = {Phys. Rev. X},
  volume = {14},
  pages = {041030},
  year = {2024}
}

@article{Yam2025,
  author = {Yam, W. K. and others},
  title = {Interconnecting quantum processors},
  journal = {npj Quantum Information},
  volume = {11},
  pages = {87},
  year = {2025}
}

@article{Vazquez2024,
  author = {Vazquez, A. C. and Tornow, C. and Rist{\`e}, D. and Woerner, S. and Takita, M. and Egger, D.},
  title = {Scaling quantum computation by connecting processors},
  journal = {Nature},
  volume = {636},
  pages = {75},
  year = {2024}
}

@article{Wu2025,
  author = {Wu, B. and others},
  title = {Distributed quantum computing with modular architectures},
  journal = {arXiv:2506.01657},
  year = {2025}
}

@article{Dalton2025,
  author = {Dalton, K. and others},
  title = {Distributed quantum processing with modular devices},
  journal = {PRX Quantum},
  volume = {6},
  pages = {040365},
  year = {2025}
}

@article{Jeng2025,
  author = {Jeng, M. J. and others},
  title = {Quantum interconnects for modular processors},
  journal = {arXiv:2501.08478},
  year = {2025}
}

@article{HarrowLow2009,
  author = {Harrow, A. W. and Low, R. A.},
  title = {Random quantum circuits are approximate 2-designs},
  journal = {Commun. Math. Phys.},
  volume = {291},
  pages = {257--302},
  year = {2009}
}

@article{BrownViola2010,
  author = {Brown, W. G. and Viola, L.},
  title = {Convergence rates for arbitrary statistical moments of random quantum circuits},
  journal = {Phys. Rev. Lett.},
  volume = {104},
  pages = {250501},
  year = {2010}
}

@article{Haferkamp2022,
  author = {Haferkamp, J.},
  title = {Random quantum circuits are approximate unitary designs},
  journal = {Quantum},
  volume = {6},
  pages = {795},
  year = {2022}
}

@article{AndresMartinez2024,
  author = {Andres-Martinez, P. and others},
  title = {Distributed quantum computing and circuit cutting methods},
  journal = {Quantum Sci. Technol.},
  volume = {9},
  pages = {045021},
  year = {2024}
}

@article{DeBone2024,
  author = {de Bone, S. and M{\"o}ller, P. and Bradley, C. and Taminiau, T. and Elkouss, D.},
  title = {Distributed quantum computing with networked processors},
  journal = {AVS Quantum Sci.},
  volume = {6},
  pages = {033801},
  year = {2024}
}

@article{AndresMartinezHeunen2019,
  author = {Andr{\'e}s-Mart{\'i}nez, P. and Heunen, C.},
  title = {Automated distribution of quantum circuits via hypergraph partitioning},
  journal = {Phys. Rev. A},
  volume = {100},
  pages = {032308},
  year = {2019}
}

@article{Rached2025,
  author = {Rached, B. and Lopez-Agudo, I. and Rodrigo, S. and Bandic, M. and Garcia-Saez, A. and Feld, S.},
  title = {Inter-core traffic metrics for distributed quantum circuits},
  journal = {IEEE Access},
  volume = {13},
  pages = {113236},
  year = {2025}
}

@article{Weinstein2008,
  author = {Weinstein, Y. S. and Brown, W. G. and Viola, L.},
  title = {Parameters of pseudo-random quantum circuits},
  journal = {Phys. Rev. A},
  volume = {78},
  pages = {052332},
  year = {2008}
}

@article{Domingo2022,
  author = {Domingo, L. and Carlo, G. G. and Borondo, F.},
  title = {Optimal quantum reservoir computing for the noisy intermediate-scale quantum era},
  journal = {Phys. Rev. E},
  volume = {106},
  pages = {L043301},
  year = {2022},
  doi = {10.1103/PhysRevE.106.L043301}
}

@article{Domingo2023,
  author = {Domingo, L. and Borondo, F. and Carlo, G. G.},
  title = {Taking advantage of noise in quantum reservoir computing},
  journal = {Sci. Rep.},
  volume = {13},
  pages = {8790},
  year = {2023},
  doi = {10.1038/s41598-023-35461-5}
}

@article{Bertrand2025,
  author = {Bertrand, Corentin and Besserve, Pauline and Ferrero, Michel and Ayral, Thomas},
  title = {Turning qubit noise into an advantage: Automatic state preparation and long-time dynamics for impurity models on quantum computers},
  journal = {Phys. Rev. B},
  volume = {111},
  pages = {245159},
  year = {2025},
  doi = {10.1103/g1ky-4zd7}
}

@article{Ding2026,
  author = {Ding, Zhiyan and Dong, Yulong and Tong, Yu and Lin, Lin},
  title = {Robust ground-state energy estimation under depolarizing noise},
  journal = {APL Comput. Phys.},
  volume = {2},
  pages = {026111},
  year = {2026},
  doi = {10.1063/5.0330167}
}

@article{Trivedi2024,
  author = {Trivedi, Rahul and Franco Rubio, Adrian and Cirac, J. Ignacio},
  title = {Quantum advantage and stability to errors in analogue quantum simulators},
  journal = {Nat. Commun.},
  volume = {15},
  pages = {6507},
  year = {2024},
  doi = {10.1038/s41467-024-50750-x}
}

@article{Kashyap2025,
  author = {Kashyap, Vikram and Styliaris, Georgios and Mouradian, Sara and Cirac, J. Ignacio and Trivedi, Rahul},
  title = {Accuracy guarantees and quantum advantage in analog open quantum simulation with and without noise},
  journal = {Phys. Rev. X},
  volume = {15},
  pages = {021017},
  year = {2025},
  doi = {10.1103/PhysRevX.15.021017}
}

@misc{Molpeceres2026,
  author = {Molpeceres, Daniel and Lu, Sirui and Cirac, J. Ignacio and Kraus, Barbara},
  title = {Benchmark of quantum algorithms for ground state preparation in the presence of noise},
  eprint = {2606.20551},
  archivePrefix = {arXiv},
  primaryClass = {quant-ph},
  year = {2026},
  doi = {10.48550/arXiv.2606.20551}
}

@article{MontesPRA2026,
  author = {Montes, J. and Borondo, F. and Carlo, G. G.},
  title = {Universal configuration for optimizing randomness generation in variational distributed quantum circuits},
  journal = {Phys. Rev. A},
  volume = {113},
  pages = {032610},
  year = {2026},
  doi = {10.1103/pwz2-yg3f}
}

@book{NielsenChuang2010,
  author = {Nielsen, M. A. and Chuang, I. L.},
  title = {Quantum Computation and Quantum Information},
  publisher = {Cambridge University Press},
  address = {Cambridge},
  year = {2010}
}

@article{Mangini2022,
  author = {Mangini, S. and others},
  title = {Qubit noise deconvolution},
  journal = {EPJ Quantum Technology},
  volume = {9},
  pages = {29},
  year = {2022}
}

@article{Brandao2016,
  author = {Brand{\~a}o, F. G. S. L. and Harrow, A. W. and Horodecki, M.},
  title = {Local random quantum circuits are approximate polynomial-designs},
  journal = {Commun. Math. Phys.},
  volume = {346},
  pages = {397--434},
  year = {2016}
}

\end{document}